\documentclass[aps,floats,prb,superscriptaddress,twocolumn]{revtex4}

\usepackage{amsfonts,amsmath}
\usepackage{graphicx}
\usepackage{color}
\usepackage{hyperref}
\usepackage{bm}

\hypersetup{colorlinks=false}

\begin{document}

\title{Super-universality of eigenchannel structures and possible optical applications}

\author{Ping Fang}
\affiliation{CAS Key Laboratory of Theoretical Physics and Institute of Theoretical Physics, Chinese Academy of Sciences, Beijing 100190, China}

\author{Chushun Tian}
\email{ct@itp.ac.cn}
\affiliation{CAS Key Laboratory of Theoretical Physics and Institute of Theoretical Physics, Chinese Academy of Sciences, Beijing 100190, China}

\author{Liyi Zhao}
\affiliation{Institute for Advanced Study, Tsinghua University, Beijing 100084,China}

\author{Yury P. Bliokh}
\affiliation{Department of Physics, Technion-Israel Institute of Technology, Haifa 32000, Israel}
\affiliation{CAS Key Laboratory of Theoretical Physics and Institute of Theoretical Physics, Chinese Academy of Sciences, Beijing 100190, China}
\affiliation{ Theoretical Quantum Physics Laboratory, RIKEN Cluster for Pioneering Research, Wako-shi, Saitama 351-0198, Japan }

\author{Valentin Freilikher}
\affiliation{Department of Physics, Bar-Ilan University, Ramat-Gan 52900, Israel}
\affiliation{CAS Key Laboratory of Theoretical Physics and Institute of Theoretical Physics, Chinese Academy of Sciences, Beijing 100190, China}
\affiliation{ Theoretical Quantum Physics Laboratory, RIKEN Cluster for Pioneering Research, Wako-shi, Saitama 351-0198, Japan }

\author{Franco Nori}
\affiliation{ Theoretical Quantum Physics Laboratory, RIKEN Cluster for Pioneering Research, Wako-shi, Saitama 351-0198, Japan }
\affiliation{Department of Physics, University of Michigan, Ann Arbor, Michigan 48109-1040, USA}

\date{\today}

\begin{abstract}
The propagation of waves through transmission eigenchannels in complex media is emerging as a new frontier of condensed matter and wave physics. A crucial step towards constructing a complete theory of eigenchannels is to demonstrate their spatial structure in any dimension and their wave-coherence nature. Here, we show a surprising result in this direction. Specifically, we find that as the width of diffusive samples increases transforming from quasi one-dimensional ($1$D) to two-dimensional ($2$D) geometry, notwithstanding the dramatic changes in the transverse (with respect to the direction of propagation) intensity distribution of waves propagating in such channels, the dependence of intensity on the longitudinal coordinate does not change and is given by the same analytical expression as that for quasi-$1$D. Furthermore, with a minimal modification, the expression describes also the spatial structures of localized resonances in strictly $1$D random systems. It is thus suggested that the underlying physics of eigenchannels might include super-universal key ingredients: they are not only universal with respect to the disorder ensemble and the dimension, but also of $1$D nature and closely related to the resonances. Our findings open up a way to tailor the spatial energy density distribution in opaque materials.

\end{abstract}

\maketitle

\section{Introduction}

\label{sec:introduction}

An unprecedented degree of control reached in experiments on classical waves
is turning the dream of understanding and controlling wave propagation in
complex media into reality \cite{Rotter17}. Central to many ongoing research
activities is the concept of transmission eigenchannel \cite{Mosk08,Choi12,Choi11,Mosk12,Genack12,Tian15,Cao15,Lagendijk16,Cao16,Yamilov16,Yilmaz16,Cao17,Lagendijk18,Cao18,Tian18}
(abbreviated as eigenchannel herefater). Loosely speaking, the eigenchannel
refers to a specific wave field, which is excited by the input waveform
corresponding to the right-singular vector \cite%
{Dorokhov82,Dorokhov84,Mello88} of the transmission matrix (TM) $\boldsymbol{%
t}$. When a wave is launched into a complex medium it is decomposed into a
number of ``partial waves'', each of which propagates along an eigenchannel
and whose superposition gives the field distribution excited by the incoming
wave. Thus, in contrast to the TM, which treats media as a black box and has
been well studied \cite{Beenakker97}, eigenchannels are much less explored, in spite of the fact that these  provide rich information
about the properties of wave propagation in the interior of the media. The understanding of the spatial
structures of these channels can provide a basis for both  fundamentals
and  applications of wave physics in complex media.

So far, the emphasis has been placed on the structures of eigenchannels in
quasi $1$D disordered media \cite%
{Genack12,Tian15,Cao15,Lagendijk16,Cao16,Yamilov16}. Yet, measurements of
the high-dimensional spatial resolution of eigenchannels have been
within  experimental reach very recently \cite{Lagendijk18,Cao18}. An intriguing localization structure of eigenchannels in the transverse direction has been observed in both real and numerical experiments for a very wide $2$D diffusive slab \cite{Cao18,Tian18}. In addition,  numerical
results \cite{Tian18} have suggested that in this kind of special high-dimensional media, even in a single disorder configuration,
the eigenchannel structure can carry some
universalities that embrace quasi $1$D eigenchannels as well. Here we study the evolution of the eigenchannels in the crossover from low to higher dimension, which so far has not been explored.  This not only provides a new angle for the fundamentals of wave propagation in disordered media, but is practically important for both experiments and applications of the eigenchannel structure in higher
dimension.

\begin{figure}[h]
\centerline{\includegraphics[width=8.7cm] {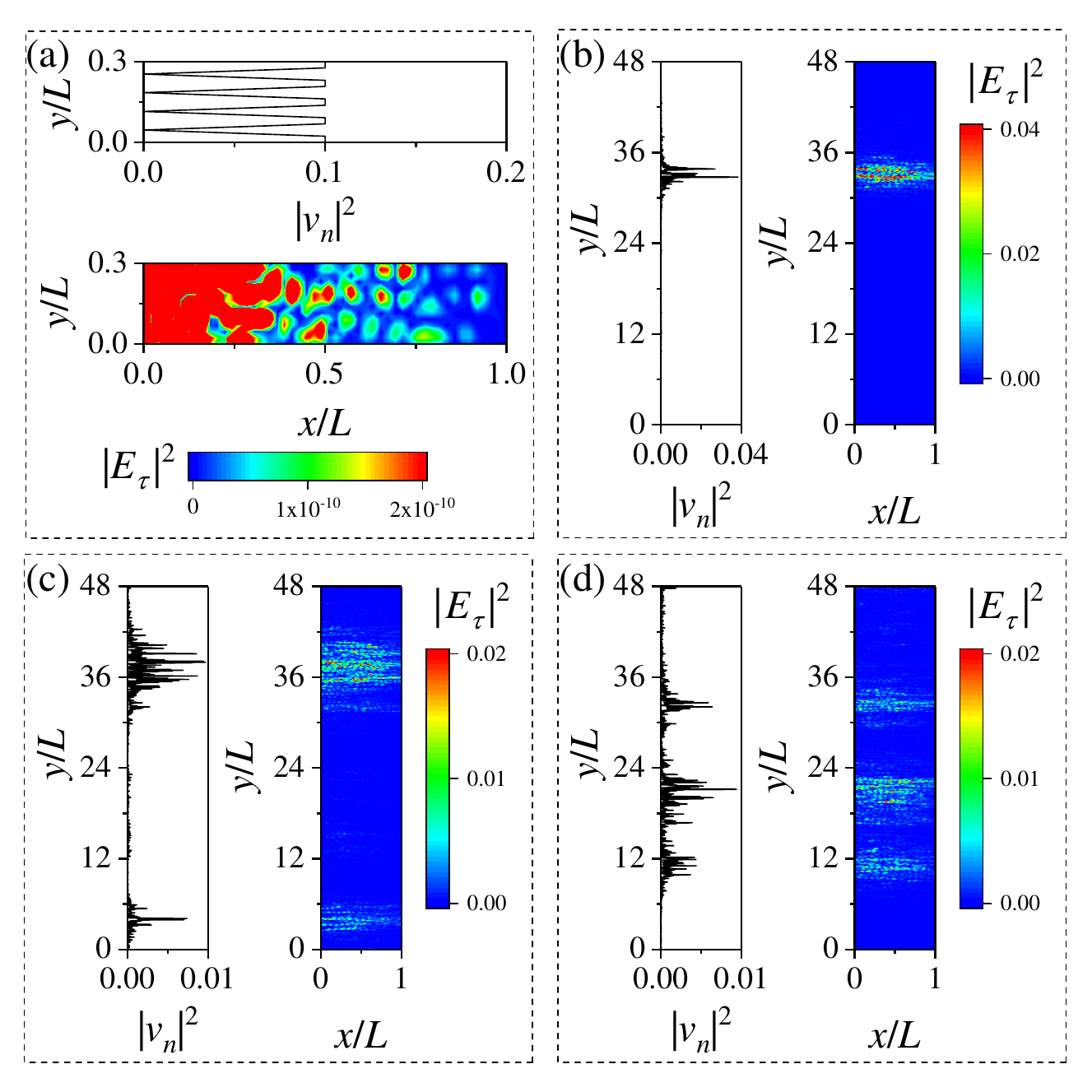}}
\caption{Simulations show that as the width $W $ increases, so that a quasi-$1$D ($N=5$%
) waveguide turns into a $2$D slab ($N=800$), the eigenchannel structure $%
|E_\protect\tau(x,y)|^2$ in a single disordered medium undergoes dramatic changes in the transverse $y$
direction. For $N=5$, the transverse structure is always declocalized (a). For $N=800$, the transverse structure exhibits very rich localization behaviors for fixed disorder configuration: for example, the structure exhibits one, two and three localization peaks in (b), (c) and (d), respectively. Moreover, the transverse structures of eigenchannels are qualitatively the same as the structures of $|v_n(y)|^2$. For all panels, the eigenvalue $\protect\tau\approx 0.6$ and $L=50$.}
\label{fig:2}
\end{figure}

Another motivation of the present work comes from a recent surprising
finding \cite{Bliokh15} regarding a seemingly unrelated object, the
resonance in layered disordered samples which, from the mathematical point
of view, are strictly $1$D systems. The resonance refers to a local maximum
in the transmittance spectrum \cite{Freilikher04}, which has a natural
connection to Anderson localization in $1$D \cite{Anderson58,Gershenshtein59}
and resonators in various systems, ranging from plasmonics to metamaterials
\cite{Freilikher03}. Despite the conceptual difference between the resonance and the eigenchannel, it was found \cite{Bliokh15} that the distribution
of resonant transmissions in the (i) $1$D \textit{Anderson localized} regime and (ii)
the transmission eigenvalues in quasi-$1$D \textit{diffusive} regime are exactly
the same, namely, the bimodal distribution \cite{Dorokhov82,Mello88}.
However, the mechanism underlying this similarity remains unclear. It is of fundamental interest to understand whether this
similarity is restricted only to transmissions, or can be extended to
spatial structures.

In this work we show that in a diffusive medium, as the width
of a sample (and consequently the number of channels) increases so that the sample crosses over from quasi $1$D to higher dimension, eigenchannels exhibit transverse structures much richer than what were found previously \cite{Cao18,Tian18}. In particular, given a disorder configuration, not only can we see the previously found \cite{Cao18,Tian18} localization structure, but also a necklace like structure which is composed of several localization peaks. Most surprisingly, notwithstanding the appearance of such diverse transverse structures in the dimension crossover, the longitudinal structure of eigenchannels, namely, the depth profile of the energy
density (integrated over the cross section), remains unaffected, and is given
by precisely the same expression as that for quasi-$1$D found in Ref.~%
\onlinecite{Tian15}. We also study the spatial structures of resonance in
strictly $1$D. We find that they have a universal analytic expression, which
is similar to that for eigenchannel structures, and the modification is
minimal. Our findings may serve as a proof of the conjecture \cite{Choi11}
of the eigenchannel structure--Fabry-Perot cavity analogy.

The remainder of this paper is organized as follows. In Sec.~\ref%
{sec:structure}, we introduce some basic concepts of eigenchannels and
resonances. In Sec.~\ref{sec:eigenchannel_structure}, we study in detail
how the eigenchannel structure in a diffusive medium evolves as the medium
crosses over from quasi-$1$D to a higher-dimensional slab geometry. To be
specific, throughout this work we focus on $2$D samples. In Sec.~\ref%
{sec:resonance_structure}, we study in detail the spatial structure of $1$D
resonances. In Sec.~\ref{sec:applications}, possible optical applications
are discussed. In Sec.~\ref{sec:discussions}, we conclude and discuss the
results.

\begin{figure}[t]
\centerline{
    \includegraphics[width=8.7cm] {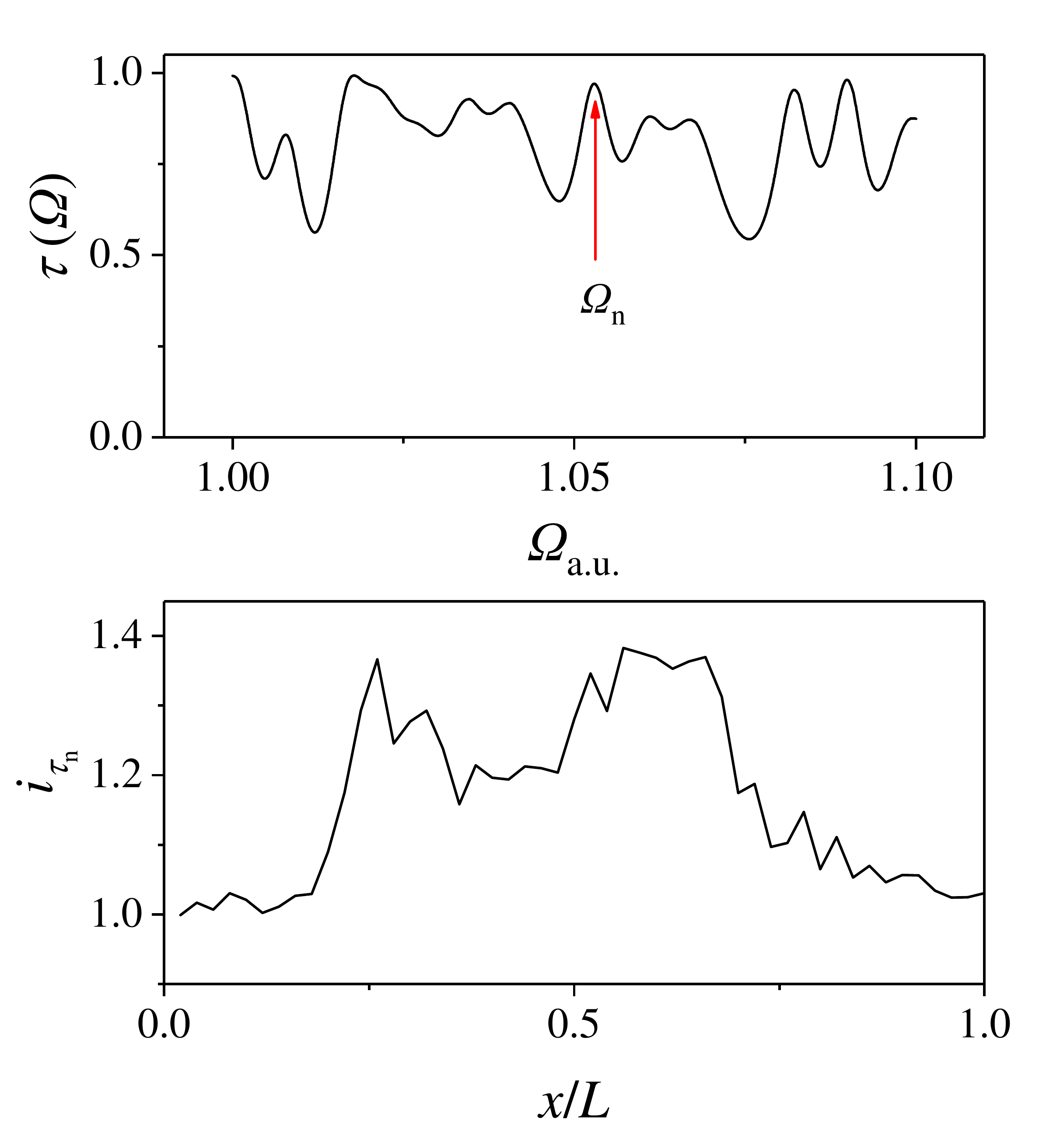}
    }
\caption{Example of a transmittance spectrum $\mathcal{T}(\Omega)$ (top) and
the resonance structure ${\tilde I}_{\mathcal{T}_n}(x)$ corresponding to the
resonant frequency $\Omega_n$ (bottom).}
\label{fig:1}
\end{figure}

\section{Eigenchannel and resonance: basic concepts}

\label{sec:structure}

To introduce the concept of eigenchannels \cite{Rotter17,Choi11,Tian15}, we
consider the transmission of a monochromatic wave (with  circular
frequency $\Omega$) through a rectangular ($0\leq x\leq L,0\leq y\leq W$)
diffusive dielectric medium bounded in the transverse ($y$) direction by
reflecting walls at $y=0$ and $y=W$. For $W\gtrsim L$ ($W\ll L$) the medium
geometry is $2$D (quasi $1$D). The wave field $E(x,y)$ satisfies the
Helmholtz equation, (the velocity of waves in the background is set to
unity.)
\begin{eqnarray}  \label{eq:6}
\left\{\partial_x^2 + \partial_y^2+\Omega^2\left[1 + \delta \epsilon (x,y)\right]\right\}E(x,y)=0,
\end{eqnarray}
where $\delta \epsilon (x,y)$ is a random function, which presents the
fluctuations of the dielectric constant inside the sample, and equals zero
at $x<0$ and $x >L$. To study the evolution of the eigenchannel structure in
the crossover from quasi-$1$D to $2$D, we increase $W$ and keep $L$, $\Omega$
and the strength of disorder fixed.

The incoming and transmitted current amplitudes are related to each other by
the transmission matrix $\boldsymbol{t}\equiv \{t_{ab}\}$, where $a,b$ label
the ideal [i.e., $\delta \epsilon (x,y)=0$] waveguide modes $\varphi_{a}(y)$%
. The matrix elements are
\begin{equation}  \label{eq:1}
t_{ab}=-i\sqrt{\tilde v_a\tilde v_b}\; \langle x=L,a|G|x^{\prime }=0,b\rangle,
\end{equation}
where $G$ is the retarded Green's function associated with Eq.~(\ref{eq:6}), and $\tilde v_a$ is the group velocity of mode $a$.

Since the matrix $\boldsymbol{t}$ is non-hermitian, we perform its
singular value decomposition, i.e., $\boldsymbol{t}=\sum_{n=1}^N \boldsymbol{u}_n \sqrt{\tau_n} \boldsymbol{v}_n^\dagger$
to find the singular value $\sqrt{\tau_n}$ and the corresponding left
(right)-singular vector $\boldsymbol{v}_n$ ($\boldsymbol{u}_n$) normalized
to unity. The input waveform $\boldsymbol{v}_n$ uniquely determines the $n$th eigenchannel, over which radiation
propagates in a random medium \cite{Rotter17,Choi11,Tian15}, and $\tau_n$
gives the transmission coefficient of the $n$th eigenchannel and is also
called the transmission eigenvalue. The total transmittance is given by $\sum_n\tau_n$.
Moreover, many statistical properties of transport through random media such
as the fluctuations and correlations of conductance and transmission may be
described in terms of the statistics of $\tau_n$ \cite%
{Tian15}.

To find the spatial structure of eigenchannels, we replace $x=L$ in Eq.~(\ref%
{eq:1}) by arbitrary $x\in [0,L)$, i.e.,
\begin{eqnarray}  \label{eq:2}
t_{ab}\rightarrow&t_{ab}(x)\equiv -i\sqrt{\tilde v_a\tilde v_b}\; \langle
x,a|G|x^{\prime }=0,b\rangle.
\end{eqnarray}
This gives the field distribution inside the medium,
\begin{equation}  \label{eq:4}
\boldsymbol{E}_{\tau_n}(x)\equiv\{E_{na}(x)\}=\boldsymbol{t}(x)\boldsymbol{v}%
_n,
\end{equation}
excited by the input field $\boldsymbol{v}_n$. Changing from the ideal waveguide mode ($\varphi_a$)
representation to the coordinate $(x,y)$ representation gives a specific $2$%
D spatial structure, namely, the energy density profile:
\begin{equation}  \label{eq:3}
|\boldsymbol{E}_{\tau_n}(x,y)|^2=\left|\sum_{a=1}^N
E_{na}(x)\varphi_a^*(y)\right|^2,
\end{equation}
which defines the $2$D eigenchannel structure associated with the
transmission eigenvalue $\tau_n$. Examples of this $2$D structure are given
in Fig.~\ref{fig:2}. Integrating Eq.~(\ref{eq:3}) over the transverse
coordinate $y$ we obtain the depth profile of the energy density of the $n$th eigenchannel,
\begin{equation}  \label{eq:5}
w_{\tau_n}(x)\equiv \int dy|\boldsymbol{E}_{\tau_n}(x,y)|^2,
\end{equation}
a key quantity to be addressed below. Note that, in the definitions of (\ref{eq:3}) and (\ref{eq:5}), the frequency $%
\Omega$ is fixed.

\begin{figure}[t]
\centerline{
    \includegraphics[width=8.7cm] {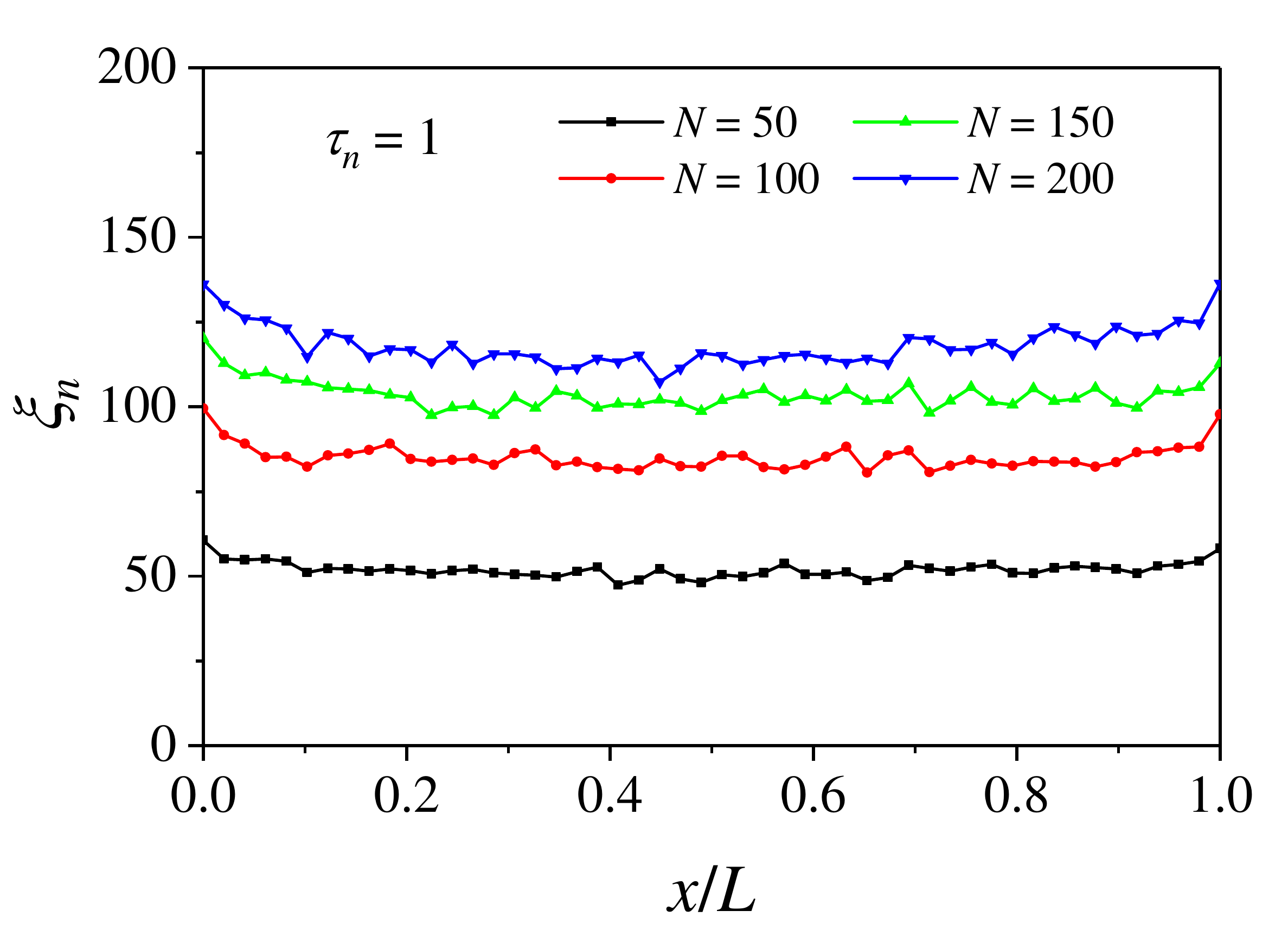}
    }
\caption{Simulation results of the $x$-dependent participation ratio $\protect\xi_n(x)$ for different values of $N$%
.}
\label{fig:9}
\end{figure}

\begin{figure*}[t]
\centerline{
    \includegraphics[width=16.40cm] {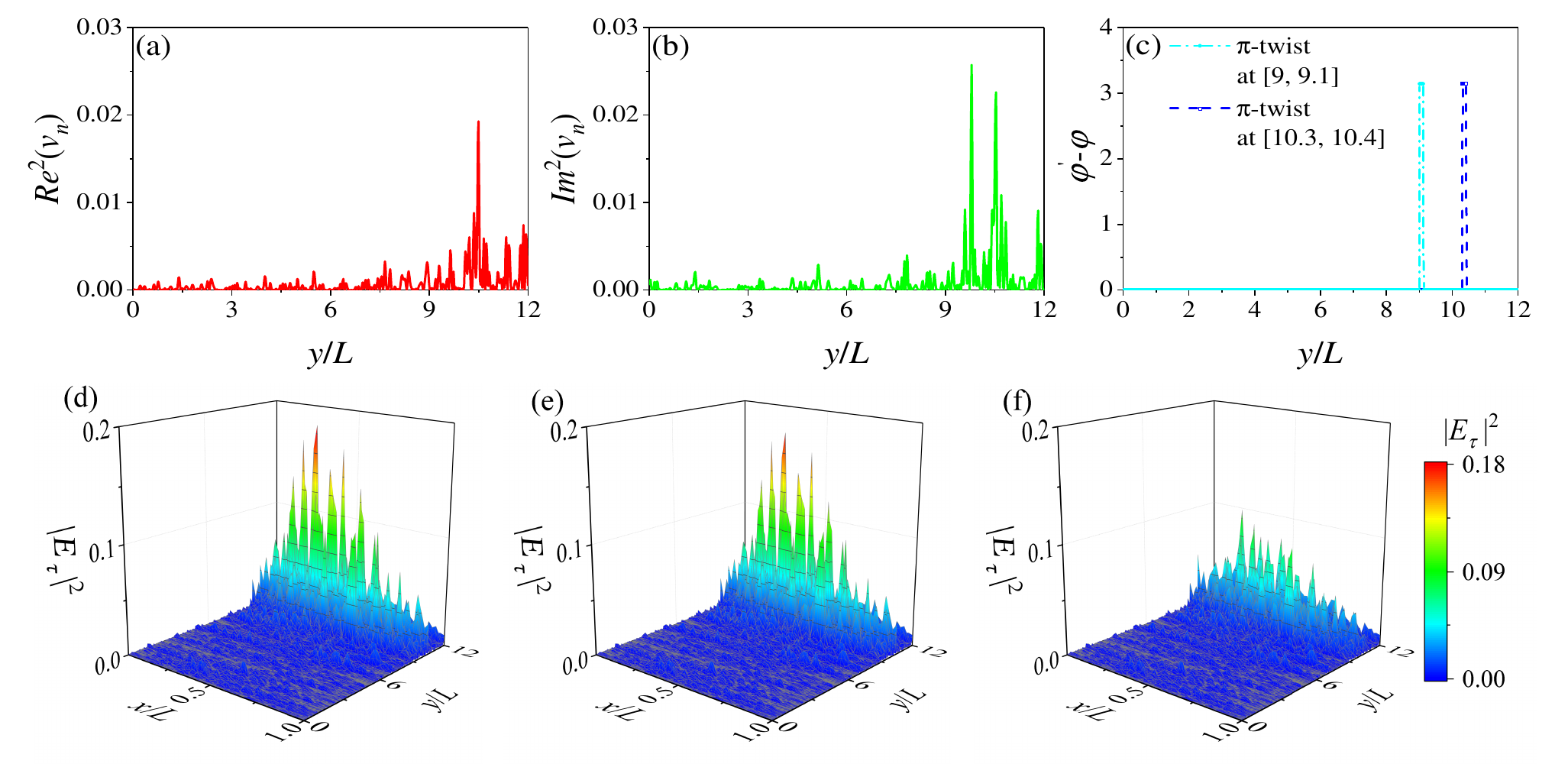}
    }
\caption{A numerical example shows that $v_n(y)$
(high transmission) is
localized approximately in the region of $9\leq y/L \leq 12$ (a,b). This
localization of $v_n(y)$ at the input edge leads to the localization of
eigenchannel structure in the $y$ direction in the interior of the medium
(d). When the phase of $v_{n}(y)$ in the region of $9.0\leq y/L \leq 9.1$ is
twisted by $\protect\pi$ (c, dotted-dashed line), the eigenchannel structure
is unaffected (e). Whereas if the $\protect\pi$-phase twist is introduced in
the region of $10.3\leq y/L \leq 10.4$ (c, dashed line), the eigenchannel
structure is significantly changed (f). $
N = 200$.
}
\label{fig:8}
\end{figure*}

\begin{figure*}[t]
\centerline{
    \includegraphics[width=16.40cm] {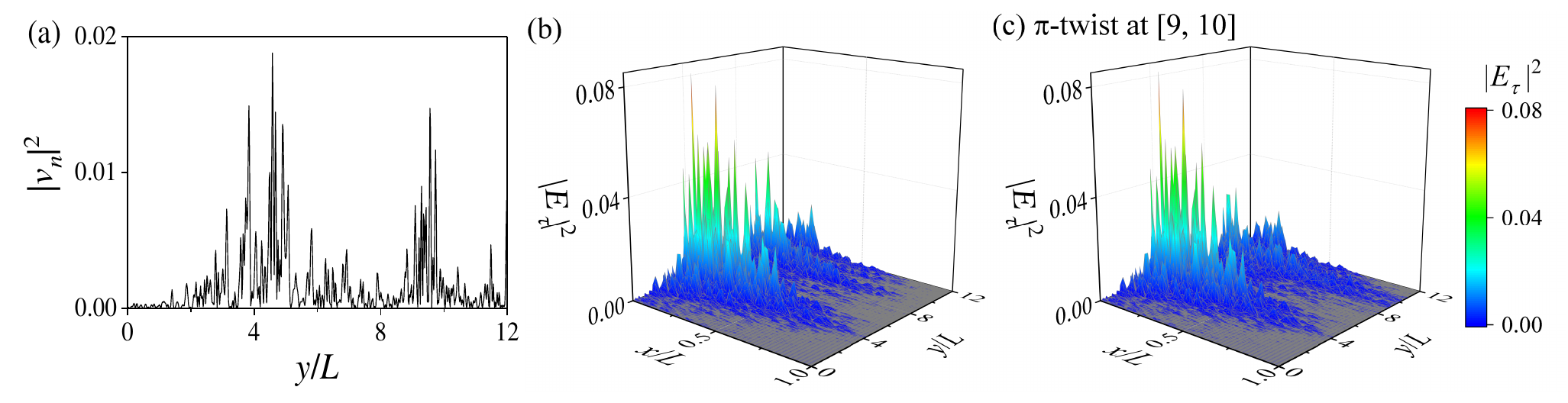}
    }
\caption{A numerical example shows that $v_n(y)$
(low transmission, $\tau_n=0.009$) exhibits two localization peaks, which are located approximately in the region of $3\leq y/L \leq 5$ and $9\leq y/L \leq 11$, respectively (a). This necklace state-like structure of $v_n(y)$ at the input edge leads to two localization peaks of
eigenchannel structure in the $y$ direction in the interior of the medium
(b). When the phase of $v_{n}(y)$ in the region of $9\leq y/L \leq 10$ is
twisted by $\protect\pi$, the localization peak of the eigenchannel structure corresponding to the localization region $9\leq y/L \leq 11$ of $v_n(y)$
is significantly changed, whereas the other peak corresponding to the localization region $3\leq y/L \leq 5$ of $v_n(y)$ is unaffected (c). $
N = 200$.}
\label{fig:10}
\end{figure*}

To proceed, we present a brief review of resonances in media with $1$D disorder
(cf.~Fig.~\ref{fig:1}). For a detailed introduction we refer to Ref.~%
\onlinecite{Freilikher03}. In the strictly $1$D case, the wave field $%
E_{\Omega}(x)$ satisfies
\begin{eqnarray}  \label{eq:7}
\left\{\partial_x^2 + \Omega^2\left[1 + \delta \epsilon (x)\right]\right\}E_{\Omega}(x)=0,
\end{eqnarray}
where $\delta \epsilon (x)$ represents the fluctuation of the dielectric
constant, and like Eq.~(\ref{eq:6}) the wave velocity at the background is set to unity. For each solution of Eq.~(\ref{eq:7}) with a given $\Omega$ there
is a specific transmittance $\mathcal{T}(\Omega)$. We define the resonance
as a local maximum $\mathcal{T}(\Omega_n)\equiv \mathcal{T}_n$ of the
transmittance spectrum $\{\mathcal{T}(\Omega)\}$, where $\Omega_n$ is the
resonant frequency. The energy density of the field at the resonant
frequency is defined as the resonance structure:
\begin{equation}  \label{eq:8}
{\tilde I}_{\mathcal{T}_n}(x)\equiv |E_{\Omega_n}(x)|^2,
\end{equation}
another key quantity to be addressed below. Importantly, contrary to the eigenchannel structure, Eq.~(\ref{eq:3}),
where $\Omega$ is fixed, to obtain the resonant
structures we need to sample $\Omega$ so that the resonances can appear.

Below we will show that although the eigenchannel in $2$D media
and the resonance in $1$D media are quite different physical
entities, their energy density spatial distributions manifest rather
surprising similarity.


\section{Eigenchannel structure in dimension crossover}

\label{sec:eigenchannel_structure}

In this section, we study numerically the evolution of eigenchannel
structures in the crossover from a quasi-$1$D ($L\gg W$) diffusive medium to
a wide ($W\gtrsim L$) $2$D diffusive slab.
We will study the energy density profiles both in
$2$D [Eqs.~(\ref{eq:4}) and (\ref{eq:3})] and in $1$D [Eq.~(\ref{eq:8})].

\subsection{Structure of right-singular vectors of the TM}

\label{sec:localization_eigenmodes_TM}

To study the eigenchannel structure given by Eq.~(\ref{eq:4}), we first
perform a numerical analysis of the transmission eigenvalue spectrum $%
\{\tau_n\}$ and the right-singular vectors $\{\boldsymbol{v}_n\}$ of the TM. We
use Eq.~(\ref{eq:6}) to simulate the wave propagation. In simulations, the
disordered medium is discretized on a square grid, with the grid spacing
being the inverse wave number in the background. The squared refractive
index at each site fluctuates independently around the background value of
unity, taking values randomly from the interval $[0.03,1.97]$. The standard recursive Green's function method \cite%
{Baranger91,MacKinnon85,Bruno05} is adopted. Specifically, we computed the
Green's function between grid points $(x^{\prime }=0,y^{\prime })$ and $%
(x=L,y^{\prime })$. From this we obtained the TM $\boldsymbol{t}$, and then numerically
performed the singular-value decomposition to obtain $\{\tau_n, \boldsymbol{v}_n, \boldsymbol{u}_n\}$.

First of all, we found that regardless of $W$, [throughout this work $%
W,L\gg \ell$ (the mean free path),] the eigenvalue density
averaged over a large ensemble of disorder configurations follows a
bimodal distribution, which was found originally for quasi $1$D samples \cite%
{Dorokhov82,Dorokhov84,Mello88} and shown later to hold for arbitrary
diffusive samples \cite{Nazarov94}.

However, as shown in Fig.~\ref{fig:2}, we found that, at a given transmission eigenvalue the spatial structure of the right-singular vector changes drastically with $W$: for small $W$ namely a quasi $1$D sample, the structure is extended (a), whereas for large $W$ namely a $2$D slab the structure is localized in a small area of the cross section, and the localization structures are very rich. Indeed, as shown in (b-d), given a disorder configuration, $|v_n(y)|^2$ can have one localization peak or several localization peaks well separated in the $y$ direction, even though these distinct structures correspond to the same
eigenvalue: the former has been found before \cite{Cao18},
while for the latter we are not aware of any reports.

\begin{figure}[t]
\centerline{
    \includegraphics[width=8.7cm] {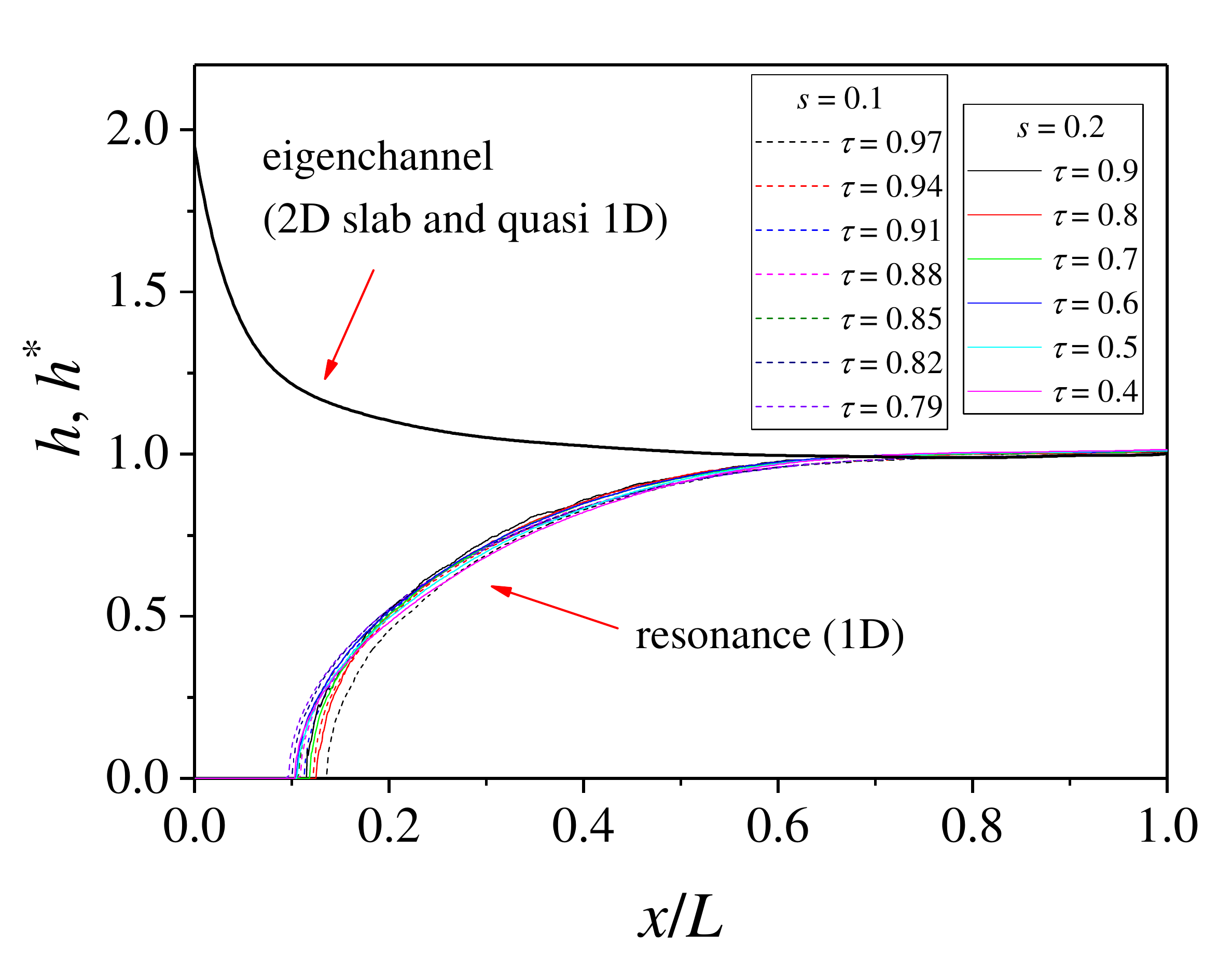}
    }
\caption{Simulations of the resonance structures in $1$D show that the $h^*$%
-function is universal with respect to both the resonant transmission $\mathcal{T%
}$ and the disorder strength $s$. This property is similar to the universality
of the $h$-function that determines the eigenchannel structures in $2$D and
quasi $1$D.}
\label{fig:3}
\end{figure}

\begin{figure}[t]
\centerline{
        \includegraphics[width=8.7cm] {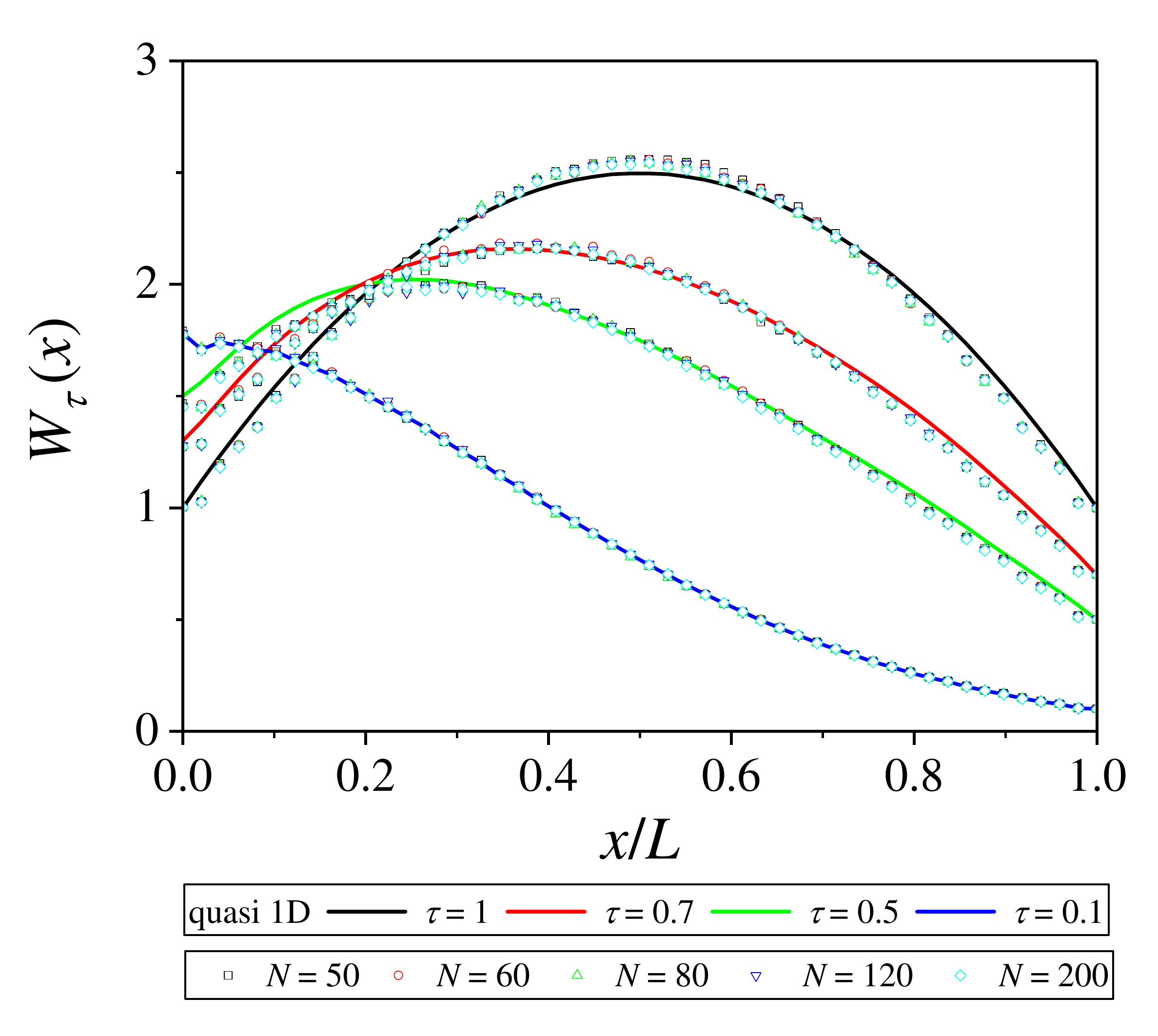}
        }
\caption{Simulations show that for $2$D slabs with different values of $N$,
the ensemble averaged depth profile $W_{\protect\tau}(x)$ (symbols) are well
described by the analytic expression given by Eqs.~(\protect\ref{eq:9})-(%
\protect\ref{eq:11}) for quasi-$1$D waveguides (black solid lines). Note that
at a given $\protect\tau$ and $x$, all symbols for distinct $N$ overlap. $L$ is fixed to be $50$ and five ratios of $W/L$ are considered,
which are $3$, $3.6$, $4.8$, $7.2$ and $12$, corresponding to $N=50$, $60$, $%
80$, $120$ and $200$, respectively.}
\label{fig:4}
\end{figure}

\subsection{Transverse structures of eigenchannels}

\label{sec:localization}

We computed the Green's function between grid points $(x^{\prime }=0,y^{\prime })
$ and $(x,y^{\prime })$, where $0\leq x\leq L$. By using Eq.~(\ref{eq:2}) we
obtained the matrix $\boldsymbol{t}(x)$. Substituting the simulation results of $\{\boldsymbol{v}_n\}
$ obtained before and $\boldsymbol{t}(x)$ into Eqs.~(\ref{eq:4}) and (\ref%
{eq:3}) we found the profile $|\boldsymbol{E}_{\tau_n}(x,y)|^2$. We repeated
the same procedures for many disorder configurations, and also for different
widths.

Figure \ref{fig:2} represents an even more surprising phenomenon occurring for very large $W$ (corresponding to $N=800$ in simulations), regardless of the transmission eigenvalues. Basically, we see that the structure of $v_n(y)$ serves as a ``skeleton'' of the eigenchannel structure: each localization peak in $|v_n(y)|^2$ triggers a localization peak in
the profile $|\boldsymbol{E}_{\tau_n}(x,y)|^2$ at arbitrary depth $x$. Thus at the cross section of arbitrary depth $x$, the transverse structure of eigenchannels is qualitatively the same as the localization structure of $|v_n(y)|^2$, i.e., if the latter has a single localization peak or exhibits a necklace structure, then so does the former, with the number of localization peaks being the same. Note that Fig.~\ref{fig:2}(b-d) correspond to the same disorder configuration and approximately the same transmission eigenvalue.

Furthermore, we introduce the $x$-dependent inverse participation ratio:
\begin{equation}  \label{eq:19}
\frac{1}{\xi_n(x)}\equiv\left\langle\frac{\int dy |\boldsymbol{E}%
_{\tau_n}(x,y)|^4}{(\int dy |\boldsymbol{E}_{\tau_n}(x,y)|^2)^2}\right\rangle
\end{equation}
associated with the field distribution $\boldsymbol{E}_{\tau_n}(x,y)$ of the
$n$th eigenchannel, where $\xi_n(x)$ characterizes the extension of the
field distribution in the $y$ direction at the penetration depth $x$, and
the average is over a number of $\boldsymbol{E}_{\tau_n}(x,y)$ corresponding
to the same singular value $\tau_n$. Figure \ref{fig:9} presents typical
numerical results of $\xi_n(x)$ for different values of $N$. From this it
is easy to see that the field distribution has the same extension in the $y$ direction for
every $x$, which is much smaller than $W$. This result provides a further evidence that the localization structure of $|v_n(y)|^2$ is maintained throughout the sample.

To understand the origin of the localization structures of eigenchannels, we first consider the case with single localization peak. We
modify the input field $\boldsymbol{v}_{n}\equiv\{v_n(y)\}$ (panels a and b
in Fig.~\ref{fig:8}) to be $\boldsymbol{v}^{\prime }_{n}\equiv\{v_n^{\prime
}(y)\}$ (panel c) in the following way. We twist by $\pi$ the phase of $%
v_{n}(y)\equiv |v_{n}(y)|e^{i\varphi(y)}$ in certain region of $y$,
\begin{eqnarray}
&&v_{n}(y)\rightarrow v^{\prime }_{n}(y)\equiv |v_{n}(y)|\exp\{i\varphi^{\prime
}(y)\}, \label{eq:16}\\
&&\varphi^{\prime }(y)=\varphi(y)+\pi\chi(y),
\label{eq:13}
\end{eqnarray}
where $\chi(y)$ takes the value of unity in the region, and otherwise of
zero. Then we let this modified input field propagate in the medium, $%
\boldsymbol{v}^{\prime }_n\rightarrow\boldsymbol{t}(x)\boldsymbol{v}^{\prime
}_n$, and compare the ensuing $2$D energy density profile with the reference
eigenchannel structure (panel d). We find that when the $\pi$-phase twist
region is away from the localization center of $v_n(y)$ (panel c,
dotted-dashed line), the resulting $2$D energy density profile is
indistinguishable from the reference eigenchannel structure (panel e). That
is, the eigenchannel structure is insensitive to  modifications. Whereas
for the changes made in the localization center (panel c, dashed
line), the ensuing energy density profile is totally different from the
reference eigenchannel structure (panel f). This shows that the localization
structures of $|\boldsymbol{E}_{\tau_n}(x,y)|^2$ are
of wave-coherence nature.

Next, we consider the case with two localization peaks. We
modify the input field $\boldsymbol{v}_{n}\equiv\{v_n(y)\}$, which has two localization peaks [Fig.~\ref{fig:10} (panel a)], in the same way as what was described by Eqs.~(\ref{eq:16}) and (\ref{eq:13}), and let the modified input field propagate in the medium. We then compare the resulting $2$D energy density  profile with the reference eigenchannel structure (panel b). Interestingly, if we perform the $\pi$-phase shift in one localization region of $|v_n(y)|^2$, then, for the ensuing $2$D energy density profile, only the peak adjacent to this localization peak of $|v_n(y)|^2$ is modified significantly, whereas the other is indistinguishable from the corresponding reference eigenchannel structure. This implies that when the transverse structure of eigenchannels if of the necklace-like shape, different localization peaks forming this necklace structure are incoherent. In addition, it provides a firm support that each localization peak in $|v_n(y)|^2$ triggers, independently, the formation of a single localization peak in the transverse structure of eigenchannels.

\subsection{Universality of eigenchannel structures in slabs}

\label{sec:universality}

Having analyzed the transverse structure of eigenchannels, we proceed to explore the longitudinal structure and to analyze its connection to the
eigenchannel structure in a quasi $1$D diffusive waveguide.

For a quasi-$1$D diffusive waveguide the ensemble average of $w_{\tau}(x)$,
denoted as $W_\tau(x)$, is given by\cite{Tian15}:
\begin{eqnarray}
W_\tau(x) = S_\tau(x)W_{\tau=1}(x),  \label{eq:9}
\end{eqnarray}
where $W_{\tau=1}(x)$ is the profile corresponding to the transparent ($%
\tau=1$) eigenchannel,
\begin{eqnarray}
W_{\tau=1}(x) = 1+\frac{\pi L x^{\prime }(1-x^{\prime })}{2\ell},\quad x^{\prime }=x/L,\quad
\label{eq:10}\\
S_\tau(x) = 2\frac{\cosh^2(h(x^{\prime })(1-x^{\prime })\phi)}{%
\cosh^2(h(x^{\prime })\phi)}-\tau,\quad \tau=\frac{1}{\cosh^2\phi},
\label{eq:11}
\end{eqnarray}
with $\phi\geq 0$. Note that $h(x^{\prime })$ increases monotonically from $h(1)=1$ as $x^{\prime }$ decreases from $1$. Its explicit form, independent of $N,\tau$, is given in Fig.~\ref{fig:3} (black solid curve).

\begin{figure}[t]
\centerline{
    \includegraphics[width=8.7cm] {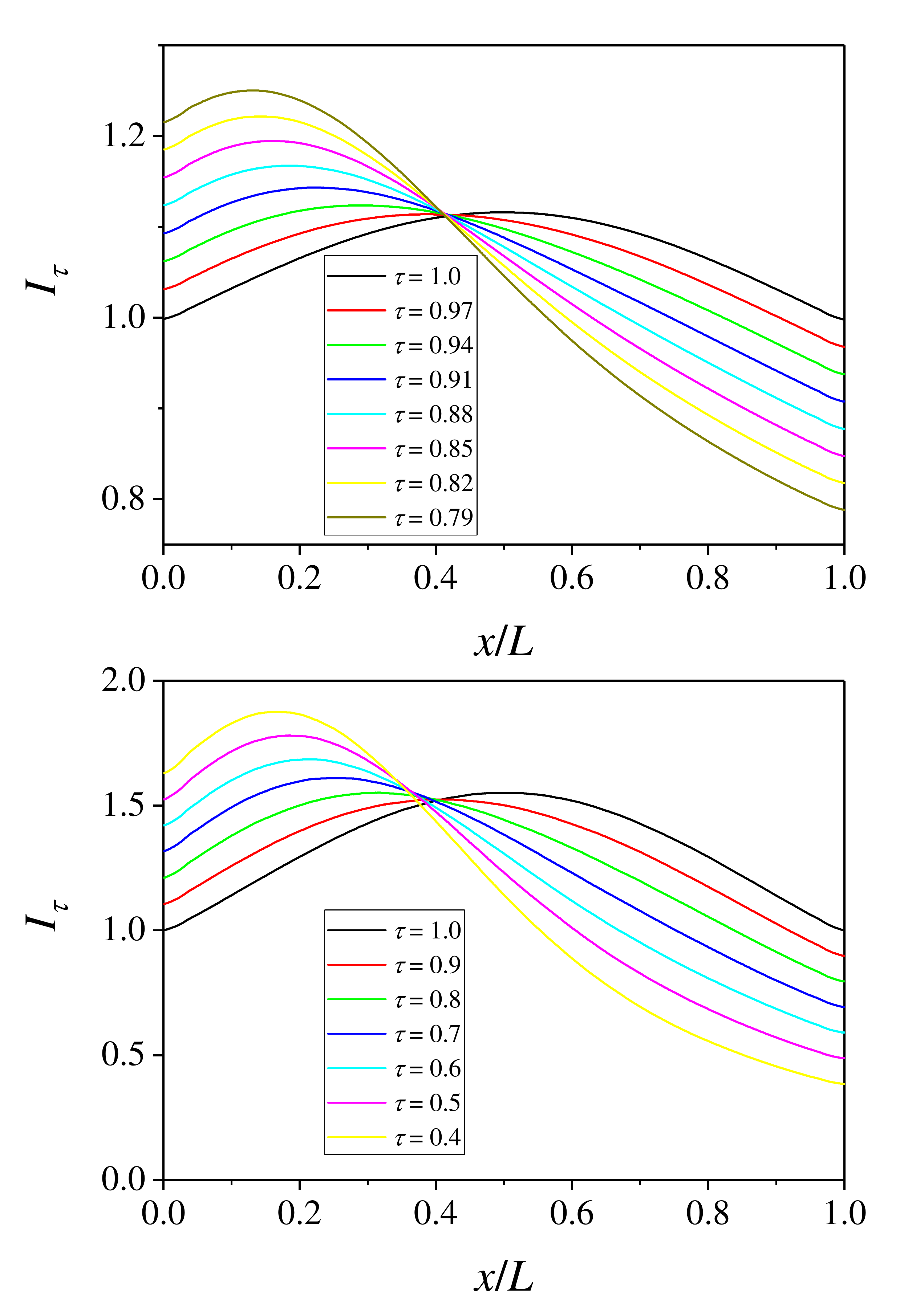}
    }
\caption{The ensemble-averaged resonance structure $I_{\mathcal{T}}(x)$ for
two different disorder strengths: $s=0.1$ (top) and $s=0.3$ (bottom), whose
corresponding localization-to-sample length ratios are $12$ and $3$,
respectively.}
\label{fig:5}
\end{figure}

\begin{figure*}[t]
\centerline{
    \includegraphics[width=16.0cm] {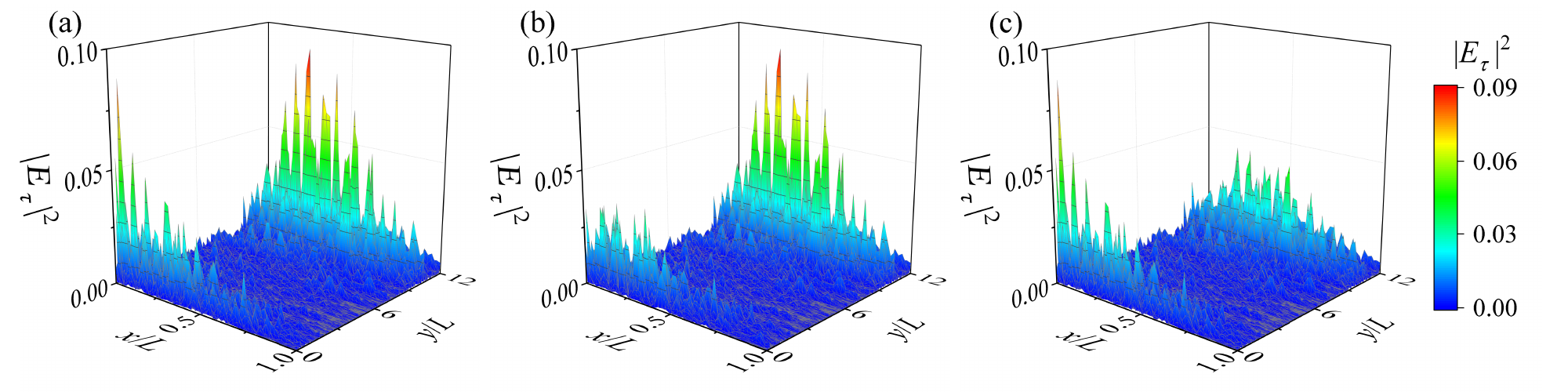}
    }
\caption{Simulations show that by shaping the input field $\boldsymbol{%
\protect\psi}_{in}$, one can realize different profiles of energy density
inside the medium. (a) The input field $\boldsymbol{\protect\psi}%
_{in}=\frac{1}{\protect\sqrt{2}}(\boldsymbol{v}_{h}+\boldsymbol{v}_{l})$,
with $\boldsymbol{v}_{h(l)}$ corresponding to the right-singular vector of
certain highly (low) transmitting eigenchannel. (b) The phase of $%
\boldsymbol{v}_{l}$ is twisted by $\protect\pi$ in the region $0.3\leq
y/L\leq 0.4$. (c) The phase of $\boldsymbol{v}_{h}$ is twisted by $\protect%
\pi$ in the region $10.3\leq y/L\leq 10.4$.}
\label{fig:7}
\end{figure*}

Now we compare the eigenchannel structure in a slab with that in a quasi-$1$%
D waveguide described by Eqs.~(\ref{eq:9}), (\ref{eq:10}) and (\ref{eq:11}).
To this end we average $2000$ profiles of $w_{\tau}(x)$ with the same or
close eigenvalues $\tau$. (Some of these eigenchannels may correspond to the
same disorder configuration.) As a result, we obtain $W_\tau(x)$ for
different values of $\tau$, shown in Fig.~\ref{fig:4}. We see, strikingly,
that for slabs with different number of $N$ (i.e., width $W$) the structures of $%
W_\tau(x)$ are in excellent agreement with $W_\tau(x)$ described by Eqs.~(%
\ref{eq:9}), (\ref{eq:10}) and (\ref{eq:11}).

\section{Resonance structure}

\label{sec:resonance_structure}

In the previous section, we have seen that in both $2$D slabs and quasi-$1$D
waveguides the ensemble averaged eigenchannel structure $W_\tau(x)$ is
described by the universal formula Eqs.~(\ref{eq:9}), (\ref{eq:10}) and (\ref%
{eq:11}). It is natural to ask whether this universality can be extended to
strictly $1$D systems, in which the transmission eigenchannel does not
exist.
Noting that the resonant transmissions have the same bimodal statistics as
the transmission eigenvalues of eigenchannels \cite{Bliokh15}, in this
section we study numerically the resonance structure ${\tilde I}_{\mathcal{T}%
_n}(x)$. It is well known \cite%
{Berezinskii73} that in strictly $1$D, there is no diffusive
regime, because the localization length is $\sim\ell$. Instead, there are only ballistic and localized regimes. We
consider the former below.

In simulations, the sample consists of $51$ scatterers separated by $50$
layers, whose thicknesses (rescaled by the inverse wave number in the
background) are randomly distributed in the interval $d_0 \pm \delta$, with $%
d_0 = 10.0$ and $\delta = 9.0$. Thus $L=50 d_0$. The scatterers are characterized by the reflection coefficients $r_i$ between the neighbouring layers ($i$ labels the scatterers.), which are
chosen randomly and independently from the interval of $(-s,s)$, with $s\in (0,1)$ governing the disorder strength. We change the frequency $\Omega$
in a narrow band centered at $\Omega_0$ and of half-width $%
5\%\times \Omega_0$, and calculate the transmittance spectrum $\mathcal{T}%
(\Omega)$ by using the standard transfer matrix approach. We also change
disorder configurations, so that for each resonant transmission $\mathcal{T}%
_n$, $5\times 10^5$ profiles of ${\tilde I}_{\mathcal{T}_n}(x)$ are
obtained. We then calculate the average of these profiles, denoted by $%
I_{\tau_n}(x)$. Finally, we repeat the numerical experiments for different values of
$s$.

Figure~\ref{fig:5} shows the simulation results of $I_\tau(x)$ for two
different values of $s$. These profiles look similar to those presented in
Fig.~\ref{fig:4}. For quantitative comparison, we compute the quantity $%
S_\tau^*(x)\equiv I_\tau(x)/I_{\tau=1}(x)$. Then, we present the function $%
S_\tau^*(x)$ in the form:
\begin{eqnarray}
S_\tau^*(x) = 2\frac{\cosh^2(h^*(x^{\prime })(1-x^{\prime })\phi)}{%
\cosh^2(h^*(x^{\prime })\phi)}-\tau,  \label{eq:12}
\end{eqnarray}
and find $h^*(x^{\prime })$ for different values of $\tau$ and $s$ from $%
S_\tau^*(x)$ calculated numerically (Fig.~\ref{fig:3}). The results are
surprising: as shown in Fig.~\ref{fig:3}, $h^*(x^{\prime })$ is a universal
function, independent of $\tau$ and $s$, which is the key feature of $%
h(x^{\prime })$ for eigenchannel structures. However, Fig.~\ref{fig:3} also
shows that the two universal functions, i.e., $h^*(x^{\prime })$ and $%
h(x^{\prime })$, are different. Therefore, allowing this minimal 
modification, the expression described by Eqs.~(\ref{eq:9}), (\ref{eq:10})
and (\ref{eq:11}) is super-universal: it applies to both the resonance
structure and the eigenchannel structure.

\section{Possible applications}

\label{sec:applications}

The universal properties of the transmission eigenchannels presented above
open up outstanding possibilities to tailor the energy distribution inside
higher-dimensional opaque materials, in particular, to concentrate energy in
different parts of a diffusive sample. In the example shown in Fig.~\ref%
{fig:7}, two eigenchannels with low ($\tau_l=0.1$) and high ($\tau_h=1$)
transmissions were exited, so that the input field had the form $\boldsymbol{%
\psi}_{in}=\frac{1}{\sqrt{2}}(\boldsymbol{v}_{h}+\boldsymbol{v}_{l})$. Here
the right-singular vector $\boldsymbol{v}_{h(l)}\equiv\{v_{h(l)}(y)\}$ of the TM
corresponds to a highly (low) transmitting eigenchannel. It is easy to see
that the energy density profile generated by this input inside the medium is
comprised of two phase coherent, but spatially separated parts. This is
because the initial transverse localization of $\boldsymbol{v}_{h(l)}$ holds
along the sample, and the integrated energy density profiles given by Eqs.~(%
\ref{eq:9}), (\ref{eq:10}) and (\ref{eq:11}) have maxima at different points
$x$ (i.e., the higher is the transmission eigenvalue, the larger is the
radiation penetration depth).


Simulations further show (Fig.~\ref{fig:7}) that it is possible, without
changing the topology of the profile, to vary the relative intensity
deposited in the two separated regions by simply modulating the phase field
of $\boldsymbol{\psi}_{in}$. For example, if we twist the phase $\varphi_l(y)
$ of $v_{l}(y)$ at points $y$ near the localization center [of $v_{l}(y)$]
by $\pi$,
\begin{eqnarray}  \label{eq:17}
&&v_{l}(y)\rightarrow v^{\prime }_{l}(y)=|v_{l}(y)|\exp\{i\varphi^{\prime
}_l(y)\},  \\
&&\varphi^{\prime }_l(y)=\varphi_l(y)+\pi \chi(y),
\end{eqnarray}
where $\chi(y)$ takes the value of unity in a region near the localization
center and otherwise is zero. For the ensuing input field $\boldsymbol{\psi}%
_{in}=\frac{1}{\sqrt{2}}(\boldsymbol{v}_{h}+\boldsymbol{v}^{\prime }_{l})$,
where $\boldsymbol{v}^{\prime }_{l}\equiv\{v^{\prime }_{l}(y)\}$, we find
that the energy density deposited in the region corresponding to the low
transmission eigenchannel is suppressed. Similarly, we can modify the input
field to suppress the energy density deposited in the region corresponding
to the high transmission eigenchannel.

\section{Conclusions and outlook}

\label{sec:discussions}

Summarizing, we have shown that in a diffusive medium, as the medium geometry crosses over from quasi-$1$D to higher dimension,
despite the transverse structure of eigenchannels (corresponding to the same transmission eigenvalues) undergo dramatic changes, i.e., from the extended  to the Anderson-like localized or necklace-like distributions, their longitudinal structure stays the same, i.e.,
the depth profile $W_\tau (x)$ of the energy density of an eigenchannel with transmission $\tau$ is always described by Eqs.~(\ref{eq:9})-(\ref{eq:11}), regardless of medium geometry. The details of the system, such as the thickness and the disorder ensemble, only enter into the ratio of $L/\ell$ in Eq.~(\ref{eq:9}). This expression is super-universal, i.e., it encompasses not only the energy distributions in diffusive eigenchannels in any dimension, but (with a minimal modification) the shape of the transmission resonances in strictly 1D random systems as well. These findings suggest that eigenchannels, which are the underpinnings of diverse diffusive wave phenomena in any dimension, might have a common origin, namely, $1$D resonances. Although the similarity between the eigenchannel structure and the Fabry-Perot cavity has been noticed already in the pioneer study of eigenchannel structures \cite{Choi11}, a comprehensive study of this phenomenon has not been carried out. The results presented above may already be helpful in further advancing the methods of focusing coherent light through scattering media by wavefront shaping. Another area of potential applications is random lasing \cite{Cao99,Wiersma08} in diffusive media. Moreover, based on previous studies \cite%
{Tian13,Bliokh15,Lagendijk16a,Tian17}, we expect that controlling the reflectivities of the edges of a sample one can tune the intensity distributions in eigenchannels, not only in quasi-1D media, but in samples of higher dimensions as well. In the future, it is desirable to explore the super-universality of eigenchannel structures in high-dimensional media, where wave interference is strong, so that Anderson localization or an Anderson localization transition occurs.

\section*{Acknowledgements}

We are grateful to A. Z. Genack for many useful discussions, and to H. Cao for informing us the preprint \onlinecite{Cao18}. C. T. is supported
by the National Natural Science Foundation of China (Grants No. 11535011 and
No. 11747601). F. N. is supported in part by the:
MURI Center for Dynamic Magneto-Optics via the
Air Force Office of Scientific Research (AFOSR) (FA9550-14-1-0040),
Army Research Office (ARO) (Grant No. W911NF-18-1-0358),
Asian Office of Aerospace Research and Development (AOARD) (Grant No. FA2386-18-1-4045),
Japan Science and Technology Agency (JST) (Q-LEAP program, ImPACT program,and CREST Grant No. JPMJCR1676),
Japan Society for the Promotion of Science (JSPS) (JSPS-RFBR Grant No. 17-52-50023, and JSPS-FWO Grant No. VS.059.18N),
RIKEN-AIST Challenge Research Fund, and the
John Templeton Foundation.


\begin{thebibliography}{99}
\bibitem{Rotter17} S. Rotter and S. Gigan,
{\ Rev. Mod. Phys.} \textbf{89}, 015005 (2017).

\bibitem{Mosk08} I. M. Vellekoop and A. P. Mosk, 
Phys. Rev. Lett. \textbf{101}, 120601 (2008).

\bibitem{Choi11} W. Choi, A. P. Mosk, Q. H. Park, and W. Choi,
{\ Phys. Rev. B} \textbf{83}, 134207 (2011).

\bibitem{Choi12} M. Kim, Y. Choi, C. Yoon, W. Choi, J. Kim, Q-H. Park,
and W. Choi, 
Nat. Photon. \textbf{6}, 581
(2012).

\bibitem{Mosk12} A. P. Mosk, A. Lagendijk, G. Lerosey, and M. Fink,
Nat. Photon. \textbf{6}, 283
(2012).

\bibitem{Genack12} M. Davy, Z. Shi, and A. Z. Genack,
Phys. Rev. B \textbf{85}, 035105 (2012).

\bibitem{Tian15} M. Davy, Z. Shi, J. Park, C. Tian, and A. Z. Genack,
{\ Nat. Commun.} \textbf{6}, 6893 (2015).

\bibitem{Cao15} S. F. Liew and H. Cao,
Opt. Express \textbf{23}, 11043 (2015).

\bibitem{Lagendijk16} O. S. Ojambati, A. P. Mosk, I. M. Vellekoop, A.
Lagendijk, and W. L. Vos,
Opt. Express \textbf{24}, 18525 (2016).

\bibitem{Cao16} R. Sarma, A. G. Yamilov, S. Petrenko, Y. Bromberg, and H.
Cao,
Phys. Rev. Lett. \textbf{117}, 086803 (2016).

\bibitem{Yamilov16} M. Koirala, R. Sarma, H. Cao, and A. Yamilov,
Phys. Rev. B \textbf{96}, 054209 (2017).

\bibitem{Cao17} C. W. Hsu, S. F. Liew, A. Goetschy, H. Cao, and A. D.
Stone, 
Nat. Phys. \textbf{13}, 497
(2017).

\bibitem{Yilmaz16} O. S. Ojambati, H. Yilmaz, A. Lagendijk, A. P. Mosk,
and W. L. Vos, 
New J. Phys. \textbf{18}, 043032 (2016).

\bibitem{Lagendijk18} P. L. Hong, O. S. Ojambati, A. Lagendijk, A. P. Mosk,
and W. L. Vos,
Optica \textbf{5}, 844 
(2018).

\bibitem{Cao18} H. Yilmaz, C. W. Hsu, A. Yamilov, and H. Cao, arXiv: 1806.01917.

\bibitem{Tian18} see Section S2 in Supplemental Materials of: P. Fang, L. Y. Zhao, and C. Tian, Phys. Rev. Lett. \textbf{%
121}, 140603 (2018).

\bibitem{Dorokhov82} O. N. Dorokhov,
Pis'ma Zh. Eksp. Teor. Fiz. \textbf{36}, 259 (1982); [JETP Lett. \textbf{36}%
, 318 (1982)].

\bibitem{Dorokhov84} O. N. Dorokhov,
Solid State Commun. \textbf{51}, 381 
(1984).

\bibitem{Mello88} P. A. Mello, P. Pereyra, and N. Kumar, Ann. Phys. (N.Y.)
\textbf{181}, 290 (1988).

\bibitem{Beenakker97} C. W. J. Beenakker, Rev. Mod. Phys. \textbf{69}, 731
(1997).

\bibitem{Bliokh15} L. Y. Zhao, C. Tian, Y. P. Bliokh, and V. Freilikher,
Phys. Rev. B \textbf{92}, 094203 (2015).

\bibitem{Freilikher04} K. Y. Bliokh, Y. P. Bliokh, and V. D. Freilikher,
J. Opt. Soc. Am. B \textbf{21}, 113 (2004).

\bibitem{Anderson58} P. W. Anderson,
Phys. Rev. \textbf{109}, 1492 (1958).

\bibitem{Gershenshtein59} M. E. Gertsenshtein and V. B. Vasil'ev, Teor.
Veroyatn. Primen. \textbf{4}, 424 (1959) [Theor. Probab. Appl. \textbf{4},
391 (1959)].

\bibitem{Freilikher03} K. Yu. Bliokh, Yu. P. Bliokh, V. Freilikher, S.
Savel'ev, and F. Nori, Rev. Mod. Phys. \textbf{80}, 1201 (2008).

\bibitem{Baranger91} H. U. Baranger, D. P. DiVincenzo, R. A. Jalabert, and
A. D. Stone,
Phys. Rev. B \textbf{44}, 10637 (1991).

\bibitem{MacKinnon85} A. MacKinnon,
Z. Phys. B \textbf{59}, 385 
(1985).

\bibitem{Bruno05} G. Metalidis and P. Bruno,
Phys. Rev. B \textbf{72}, 235304 (2005).

\bibitem{Nazarov94} Yu. V. Nazarov, Phys. Rev. Lett. \textbf{73}, 134 (1994).



\bibitem{Berezinskii73} V. L. Berezinskii, Zh. Eksp. Teor. Fiz. \textbf{65},
1251 (1973) [Sov. Phys. JETP \textbf{38}, 620 (1974)].

\bibitem{Cao99} H. Cao, Y. G. Zhao, S. T. Ho, E. W. Seelig, Q. H. Wang, and
R. P. H. Chang, 
Phys. Rev. Lett. \textbf{82}, 2278 (1999).

\bibitem{Wiersma08} D. S. Wiersma,
Nature Phys. \textbf{4}, 359 
(2008).

\bibitem{Tian13} X. J. Cheng, C. S. Tian, and A. Z. Genack, Phys. Rev. B
\textbf{88}, 094202 (2013).

\bibitem{Lagendijk16a} D. Akbulut, T. Strudley, J. Bertolotti, E. P. A. M.
Bakkers, A. Lagendijk, O. L. Muskens, W. L. Vos, and A. P. Mosk,
Phys. Rev. A \textbf{94}, 043817 (2016).

\bibitem{Tian17} X. Cheng, C. Tian, Z. Lowell, L. Y. Zhao, and A. Z. Genack,
Eur. Phys. J. Special Topics \textbf{226}, 1539 
(2017).
\end{thebibliography}
\end{document}